\documentclass[3p,times,twocolumn]{elsarticle}
 \biboptions{comma,sort&compress}
\usepackage{graphicx}
\usepackage{dcolumn}
\usepackage{amsmath}
\usepackage{here}
\usepackage{ecrc}

\volume{00}
\usepackage{multicol}
\firstpage{1}

\journalname{Nuclear and Particle Physics Proceedings}
\runauth{Kluth, Ochs and Perez-Ramos}

\jid{nppp}
\jnltitlelogo{Nuclear and Particle Physics Proceedings}

\usepackage{amssymb}

\usepackage[figuresright]{rotating}

\begin{document}

\begin{frontmatter}

\title{
%
Dead cone effect in charm and bottom quark jets} 
 
 \cortext[cor0]{Mini-Review talk presented by R. Perez-Ramos at QCD23, 26th International Conference in QCD (10-14/07/2023,
  Montpellier - FR). }

 \author[label1]{Stephan Kluth
 \corref{cor1} 
 }
   \address[label1]{Max-Planck-Institut f\"ur Physik, F\"ohringer Ring 6, 80805 M\"unchen, Germany}
\ead{skluth@mpp.mpg.de}

 \author[label1]{Wolfgang Ochs
 \corref{cor1} 
 }
\ead{ww.ochs@gmx.de}
 \author[label3]{Redamy Perez-Ramos
 \corref{cor3} 
 }
   \address[label3]{DRII-IPSA, Bis, 63 Boulevard de Brandebourg, 94200 Ivry-sur-Seine, France\\
and\\
Laboratoire de Physique Th\'eorique et Hautes Energies (LPTHE), UMR 7589,\\ Sorbonne Universit\'e et CNRS, 4 place Jussieu, 75252 Paris Cedex 05, France
}
\ead{redamy.perez-ramos@ipsa.fr}

\pagestyle{myheadings}
\begin{abstract}
\noindent
The evolution of a heavy quark initiated jet is mainly ruled by gluon bremsstrahlung. As a consequence of the dead-cone effect, this radiation is suppressed in the forward direction at angles smaller than that proportional to the heavy quark mass $M_Q$, i.e. $\Theta_0=M_Q/E_Q$ at energy $E_Q$ of the primary quark. In this paper, we unveil this effect in charm and bottom quark jets using DELPHI and OPAL data from Z$^0$ boson decays in $e^+e^-$ annihilation at center of mass energy 91.2 GeV. The analysis of the reconstructed heavy quark fragmentation function in momentum space shows the strong suppression of hadrons at high momenta in such events compared to light quark fragmentation by a factor $\lesssim1/10$. The amount of this suppression is well reproduced by perturbative QCD (pQCD) within the Modified Leading Logarithmic Aproximation and the compact scheme of Local Parton Hadron Duality (MLLA-LPHD). As a new result, we obtain an almost perfect agreement between the light quark fragmentation functions expected at $W_0\propto M_Q$ from DELPHI and OPAL data with Pythia8 and shed light on the reasons for the existence of the ultra-soft gluon excess at small momentum fraction in comparison with pQCD predictions.
\begin{keyword}  Heavy quarks, QCD jets,  dead cone efffect, fragmentation function.


\end{keyword}
\end{abstract}
\end{frontmatter}
\section{Introduction}
In this paper, we study the dead cone effect in charm and bottom quark jets in QCD, the theory of strong interactions within the Standard Model of particle physics. The radiation pattern off a heavy quark  for the splitting process $Q(\bar{Q})\to Q(\bar{Q})+g$ reads~\cite{Dokshitzer:1991fc,Dokshitzer:1991fd}
\begin{equation}
  d\sigma_{Q\to Q+g} \simeq \frac{\alpha_s}{\pi}C_F\frac{d\omega}{\omega}\frac{\Theta^2d\Theta^2}{(\Theta^2+\Theta^2_0)^2},
\label{emission}
\end{equation}
where, $\Theta_0=M_Q/E_Q$ is the angular cut-off; $M_Q$ is the heavy quark mass, $E_Q$ the heavy quark initial energy, $\omega$ the emitted gluon energy, $\alpha_s$ denotes the strong coupling constant and $C_F$ the QCD colour factor at the branching vertex $Q(\bar{Q})\to Q(\bar{Q})+g$. Thus, for emission angles $\Theta<\Theta_0$, gluon radiation vanishes in the forward direction such that the region with the gluon depopulated cone around the flight direction of the heavy quark $Q$ is called ``dead cone''. For emission angles $\Theta \gg \Theta_0$, the gluon radiation pattern becomes identical to that of a light quark jet, and the same statement holds for the internal angular ordered structure of secondary gluon subjets. For light quarks in the limit $M_Q\to0$, the radiation pattern (\ref{emission}) reduces to the known double soft and collinear differential logarithmic form
\begin{equation}
  d\sigma_{q\to q+g} \simeq \frac{\alpha_s}{\pi}C_F\frac{d\omega}{\omega}\frac{d\Theta^2}{\Theta^2}.
\label{dl_emission}
\end{equation}
In previous studies, as a first consequence of the dead cone effect, a reduction of the full particle multiplicity in the heavy quark jet has been predicted. This effect has indeed been observed in~\cite{Schumm:1992xt} and in the subsequent update~\cite{Dokshitzer:2005ri} with results nearby the QCD expectation. Later, a more direct observation of the dead cone effect has been achieved by the ALICE collaboration~\cite{ALICE:2021aqk}, which has presented a relative suppression of small angle particle emission in charm quark jets by a factor of about $\sim1/2$ in agreement with Monte Carlo Event Generators (MCEG). In the most recent study \cite{Kluth:2023umf}, the dead cone effect in c- and b-quark events is put forward using data on momentum spectra from Z boson decays in $e^+e^-$ annihilation. The heavy quark fragmentation function for charged particles is reconstructed in the momentum fraction variable $x$ or $\xi=\ln(1/x)$ for the first time by removing the decays of the heavy quark hadrons. 
%
%
%
\section{Unveiling the dead cone effect in heavy quark events}
In \cite{Kluth:2023umf}, experimental data are reanalysed as function of the charged particle momenta $p$, with $x_p=2p/W$ or $\xi_p=\ln (1/x_p)$ at c.m.s.\ energy $W$. In $e^+e^-$ annihilation, we refer to the fragmentation function in one hemisphere as
\begin{equation}\label{eq:FF}
  \bar{D}(\xi,W)=\frac{1}{2}\frac{1}{\sigma_{tot}} \frac{d\sigma^h}{d\xi} (\xi,W)
\end{equation}
where $\bar D(\xi,W)=x D(x,W)$ with the inclusive $x$-distribution $D(x,W)$. In order to obtain the heavy quark fragmentation function $\bar{D}_Q^{ch}(\xi_p,W)$, $Q=c,b$, we start from the measured $\xi_p$-distribution of light hadrons in events tagged as originating from $Z\rightarrow Q{\bar Q}$ decays. This distribution also contains the charged hadrons from B-hadron or Charm-hadron decays and they have been subtracted~\cite{Kluth:2023umf}. The $\xi_p$-distributions of charged B-hadron or Charm-hadron decay products have not been measured separately and we obtained them from a MCEG program; the most recent version of Pythia8~\cite{Bierlich:2022pfr,Skands:2014pea} was used for this purpose. In the end, the $\xi_p$-distributions of charged B-hadron or Charm-hadron decay products are subtracted from the full $\xi_p$-distributions of heavy quark events in order to obtain the derived b- and c-quark fragmentation functions from DELPHI and OPAL displayed in~\cite{Kluth:2023umf}. These are to be compared with the light quark $\xi_p$-distributions so as to unveil the dead-cone effect.
\begin{figure*}[htpb!]
\begin{multicols}{2}
\hskip -1.1cm
    \includegraphics[height=9.2cm,width=9.5cm]{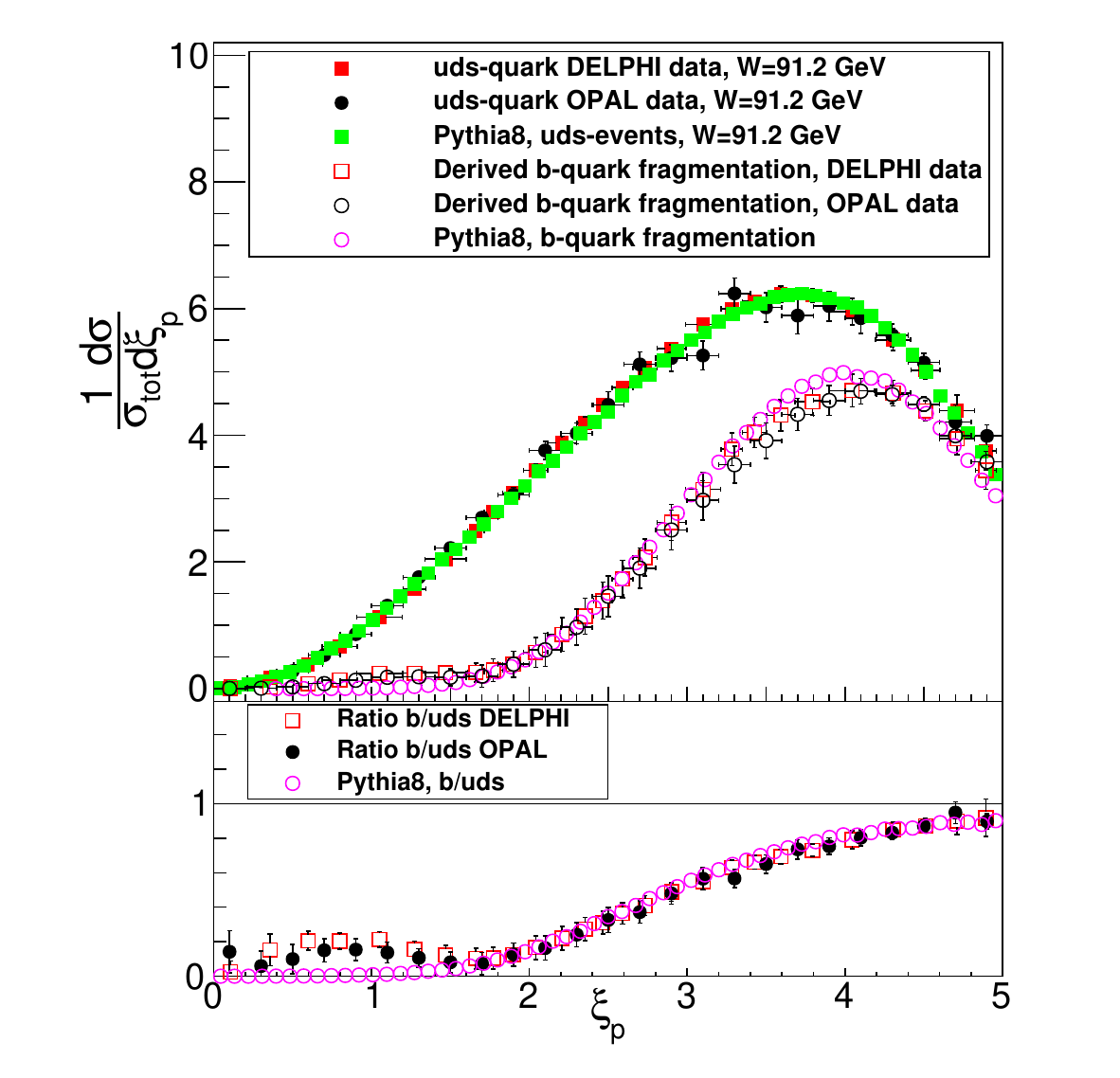}\par 
\hskip -1cm    
    \includegraphics[height=9.2cm,width=9.5cm]{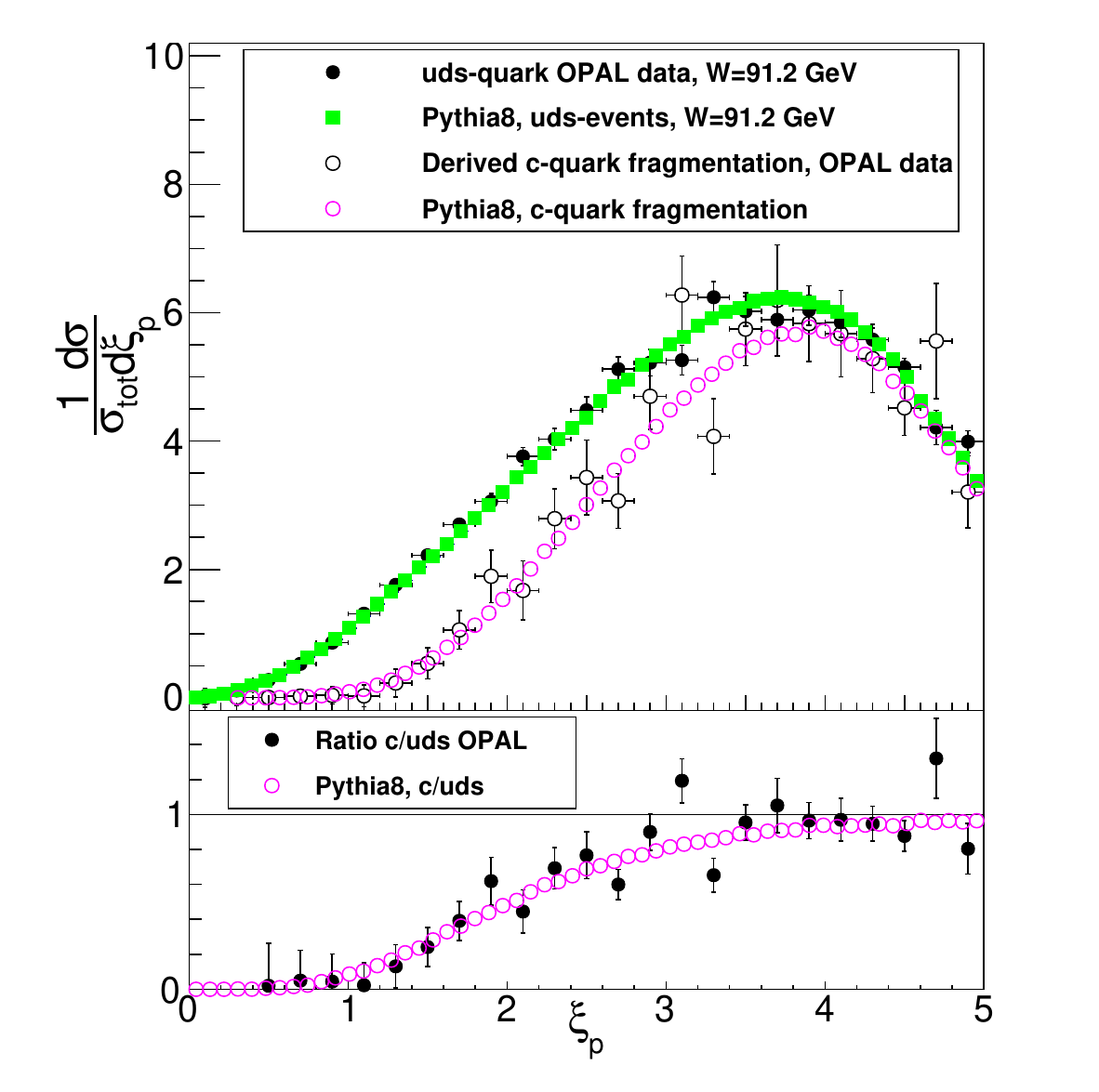}\par 
    \end{multicols}
    \vskip -0.5cm
\caption{Fragmentation function in $\xi_p= \ln (1/x_p)$ for the light uds-quarks in comparison with experimentally derived $\xi_p$-distributions for the b-quark (left panel) and c-quark (right panel) fragmentation (upper panels); ratio of the heavy quark over the light quark fragmentation functions showing the strong suppression of the heavy quark fragmentation for small $\xi_p$ (large momenta) by an order of magnitude which constitutes the dead cone effect (lower panels) in generally good agreement with MCEG Pythia8.}
\label{fig:exp-deadcone}
\end{figure*}

For this purpose, in Fig.~\ref{fig:exp-deadcone} for $\sqrt{s}=91.2$ GeV, we display the uds-quark $\xi_p$-distributions with the derived b- and c-quark $\xi_p$-distributions in order to show the increasing suppression of particles with decreasing $\xi_p$  approaching a factor of about $\sim1/10$ as compared to light quark fragmentation for particles with $\xi_p\lesssim1.6$ in b-quark jets and $\xi_p\lesssim 1$ in c-quark jets. This corresponds in momentum space to $x\gtrsim0.2$, or $9\lesssim p\;(\text{GeV/c})<45.6$ in b-quark jets and $x\gtrsim0.4$, or $18\lesssim p\;(\text{GeV/c})<45.6$ in c-quark jets. Also shown are the Pythia8 results for uds-, b- and c-quark fragmentation functions which turn in an almost perfect agreement with experimental data except for the b/uds-ratio at small $\xi_p$ where hard particle suppression at large $x\sim1$ is  most severe as compared to the derived DELPHI and OPAL data sets.
The sensitivity to the dead cone effect in the present momentum analysis is considerably increased in comparison to the presented angular analysis by ALICE~\cite{ALICE:2021aqk}. This may be related to the finite jet resolution and the difficulty to define the gluon emission angle in that analysis. 
\section{Comparison with the MLLA-LPHD scheme}

The spectrum $D_Q(x,E)$ of gluons with energy fraction $x=E_g/E$ at primary energy $E$ accompanying the $Q{\bar Q}$ pair can be treated in a similar way to that of mean multiplicities; for a review, see~\cite{Khoze:2001aa}. The difference between the heavy quark $D_Q(x,E)$ and the light quark $D_q(x,E)$ spectra due to the dead cone effect comes from the radiation of very energetic gluons at small angles $\Theta < \Theta_0$. This radiation can be considered as resulting from a Lorentz boost by the factor $\gamma= E/M_Q$ along the heavy quark direction from the corresponding radiation at lower hardness $M_Q$.

The corresponding analysis for the inclusive spectra in MLLA is not yet available at the same rigor multiplicities have been derived. An approximate equation for the inclusive $x$-spectra has been presented which reproduces the equation for multiplicities in MLLA after integration over $x$ and avoids a negative fragmentation function. The MLLA estimate has been reported as~\cite{Dokshitzer:1991fd}
\begin{equation}
  \bar D_Q(x,W)= \bar D_q(x,W) -  \bar D_q\left(\frac{x}{\langle x_Q\rangle},\sqrt{e}M_Q\right), 
\label{MLLAeq}
\end{equation}
where $\bar D(x,W)=x D(x,W)$. 
This expression, after integration over the variable $x$, reproduces the MLLA-derived multiplicities~\cite{Schumm:1992xt,Dokshitzer:2005ri}. For our comparison with the heavy quark fragmentation function $\bar D_Q(\xi,W)$ with variable $\xi$, we rewrite this relation as
\begin{equation}
  \bar D_Q(\xi,W)=\bar D_q(\xi,W) -  \bar D_q( \xi - \xi_Q,\sqrt{e}M_Q),
\label{MLLAeqxi}
\end{equation}
with $\bar D(\xi,W)=x D(x,W)$ and $\xi_Q=\ln(1/\langle x_Q\rangle)$. Again, as in the equation for multiplicities, the low energy scale is $W_0=\sqrt{e} M_Q$ with the large MLLA correction factor $\sqrt{e}\approx1.65$. Furthermore, the mean momentum fraction ${\langle x_Q\rangle}$ of the primary heavy quark $Q$ is introduced which reduces the light particle energies to $x < {\langle x_Q\rangle}$ and guarantees that the final accompanying radiation distribution stays positive over the allowed kinematic region. The shift of the $\xi$-spectrum by $\xi_Q$ corresponds to an MLLA correction of ${\cal O}(\sqrt{\alpha_s})$ as can be seen by a Taylor expansion of $\bar D_q(\xi-\xi_Q,\sqrt{e}M_Q)$ in $\xi_Q$ at  $\xi_Q=0$. The eq.~\eqref{MLLAeqxi} represents an approximation that does not work well at small $\xi$ since the shifted contribution $\bar D_q(\xi-\xi_Q,W_0)$ has to vanish for $\xi<\xi_Q$. Comparisons of these predictions with experiment should take these limitations into account.

If ${\langle x_Q\rangle}$ is taken from experiment, the heavy quark fragmentation function at c.m.s.\ energy $W$ can be obtained by the relation eq.~\eqref{MLLAeqxi} from the light quark fragmentation functions at energies $W$ and $W_0$ in absolute normalisation.

As numerical values of these parameters, we take for b-quarks, $W_0=8.0$~GeV\footnote{This value corresponds to a b-quark pole mass $M_b=4.85\pm0.15$ which is consistent with the most recent world average pole mass $M_b=4.78\pm0.06$~\cite{ParticleDataGroup:2022pth}}~\cite{Dokshitzer:2005ri} and the experimental evaluation $\langle x_b \rangle=0.7092\pm0.0025$~\cite{DELPHI:2011aa}. For c-quarks, we use $W_0=2.7$~GeV~\cite{Dokshitzer:2005ri} and the experimental value $\langle x_c\rangle=0.495\pm0.006$~\cite{Baines:2006uw}. This yields the shift parameters
\begin{equation}
  \xi_c=0.70, \ \ \  \xi_b=0.36.
\label{xicbvalues}
\end{equation}
These numbers are also consistent with the results in~\cite{Khoze:1996dn}, based on calculations for the heavy quark $x$-spectra in~\cite{Dokshitzer:1995ev}.

We first probe the MLLA expectation eq.~\eqref{MLLAeqxi} by inserting for $\bar D(\xi,W)$ the experimentally observed distributions in $\xi_p=\ln{1/x_p}$ at the respective energies $W$.
At the low energy $W_0=2.7$~GeV for the c-quark fragmentation, we insert the $\xi_p$-distribution data obtained by the BES collaboration at the nearby energy 2.6~GeV~\cite{BES:2003xdf}. There are no data nearby $W_0=8.0$ GeV for the b-quark fragmentation and therefore, we obtain the corresponding $\xi_p$-distribution from the interpolation between two neighbouring energies, also a correction for charm production has been applied (see \cite{Kluth:2023umf}). The $\xi_p$-distribution at $W_0=8.0$~GeV, so obtained and shifted by $\xi_b=0.36$ according to eq.~\eqref{MLLAeqxi}, i.e.\ $\bar D_q(\xi-\xi_b,W_0)$. From their difference, according to eq.~\eqref{MLLAeqxi}, one obtains the MLLA predictions for the b- and c- quark distributions $\bar D_b(\xi_p,W)$ and $\bar D_c(\xi_p,W)$ where the error bars shown include the systematic errors. 
\begin{figure*}[htpb!]
\begin{multicols}{2}
\hskip -1.1cm
\includegraphics[height=8.0cm,width=9.5cm]{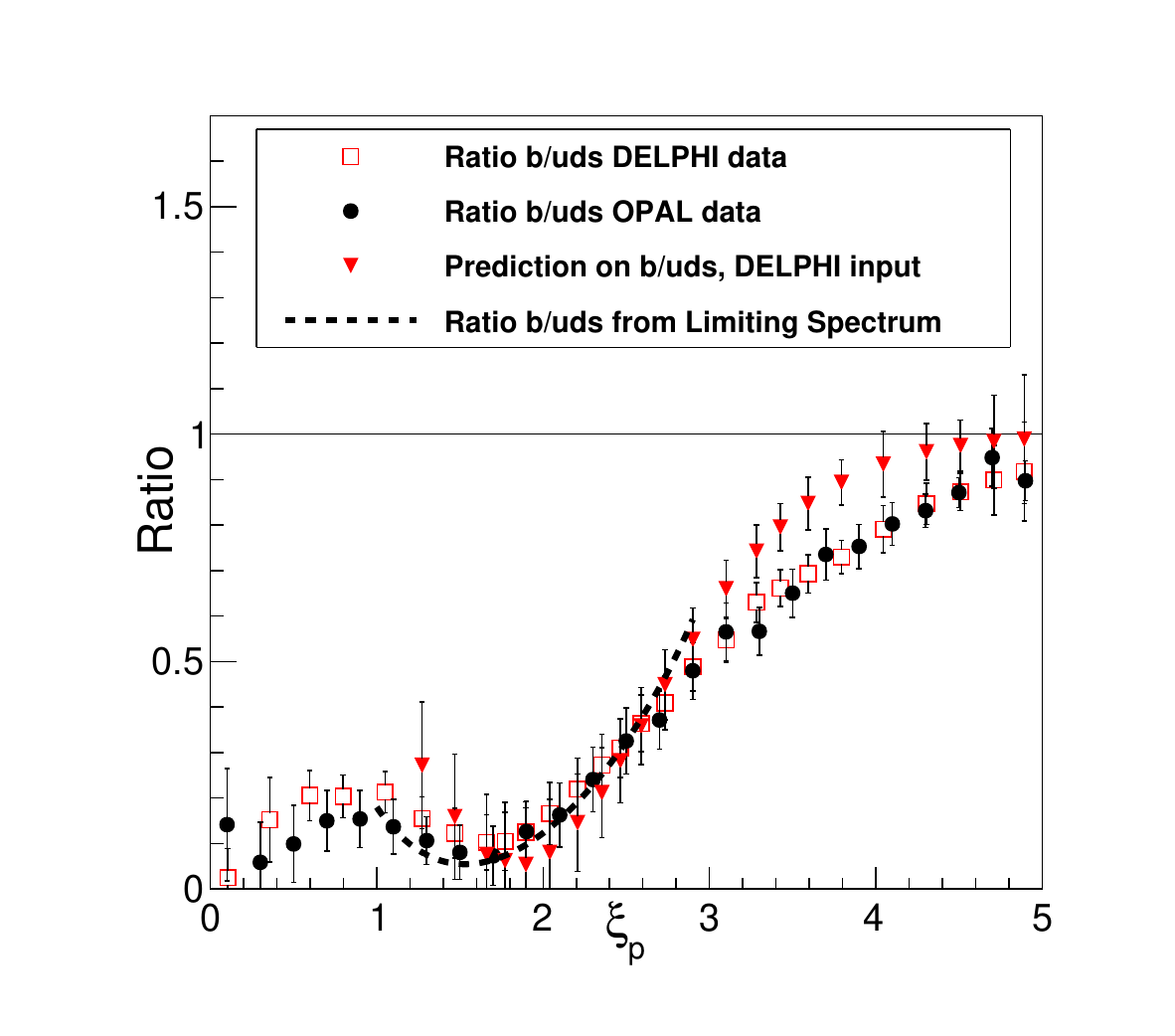}\par 
\hskip -1cm    
\includegraphics[height=8.0cm,width=9.5cm]{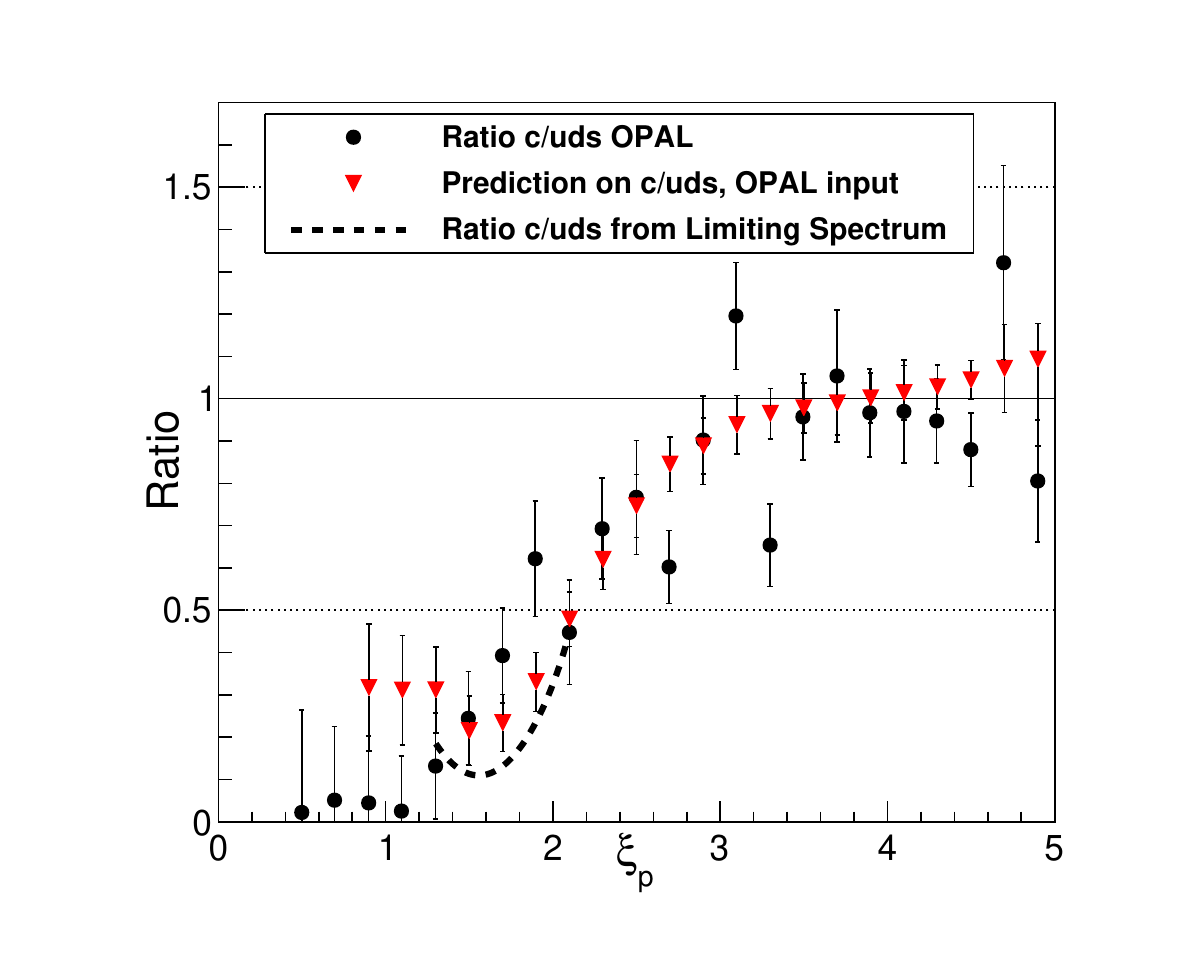}\par 
\end{multicols}
\vskip -0.5cm
\caption{Ratio of the heavy b-quark over the light uds-quark fragmentation functions (left panel) and the corresponding ratio for c-quark (right panel) together with the MLLA expectations based on comparison with experimental data, with Limiting Spectrum distributions and with Pythia8.}
\label{fig:exp-deadcone_ratios}
\end{figure*}

In Fig.~\ref{fig:exp-deadcone_ratios} we display again the ratio of heavy quark over light quark fragmentation functions in comparison with the MLLA prediction eq.~\eqref{MLLAeqxi} using either data input or Limiting Spectrum fits. These ratios are correctly reproduced for b-quarks in the central region $1\lesssim\xi_p\lesssim 3$. For c-quarks the predictions from data are generally compatible with the measurements within the large errors whereas the predictions from the Limiting Spectrum agree with the data within the accessible region $1\lesssim\xi_p\lesssim 2$. 
Below $\xi_p \sim 1$ ($x\gtrsim 0.4$) the Limiting Spectrum ratios are rising again, because of a mismatch in the lower limit in $\xi_p$ for the limiting spectrum at 91.2~GeV and the shifted one at the low energy $W_0$ (i.e.\ at 2.6 or 8~GeV), therefore we excluded those results from the figure (see also next section). The amount of suppression from the dead cone effect is correctly reproduced for both heavy quark fragmentation processes.

\section{On the reconstruction of the uds-quark FF at $W_0=\sqrt{e}M_Q$: comparison with Pythia8}
In the figure~\ref{fig:exp-deadcone_ratios} for the ratios, it is shown that the MLLA expectation eq.~\eqref{MLLAeqxi} quantitatively predicts the suppression of particle production in the central region around the maximum of the $\xi_p$-distribution at the lower mass scale $W_0$. For the b-quark, however, there is a major surplus of particles at large $\xi_p$ beyond expectation and a smaller excess at small $\xi_p$. To quantify these effects more clearly, we investigate the difference between the heavy and light quark fragmentation functions at $W=91.2$~GeV which, according to the MLLA expectation eq.~\eqref{MLLAeqxi}, should just yield the expected light quark $\xi_p$-distribution at the lower mass scale $W_0$. This will clarify how the predicted $\xi_p$- distribution at the low energy $W_0$ deviates from the observed one. 
\begin{figure*}[htpb!]
\begin{multicols}{2}
\hskip -1.1cm
\includegraphics[height=9.5cm,width=9.5cm]{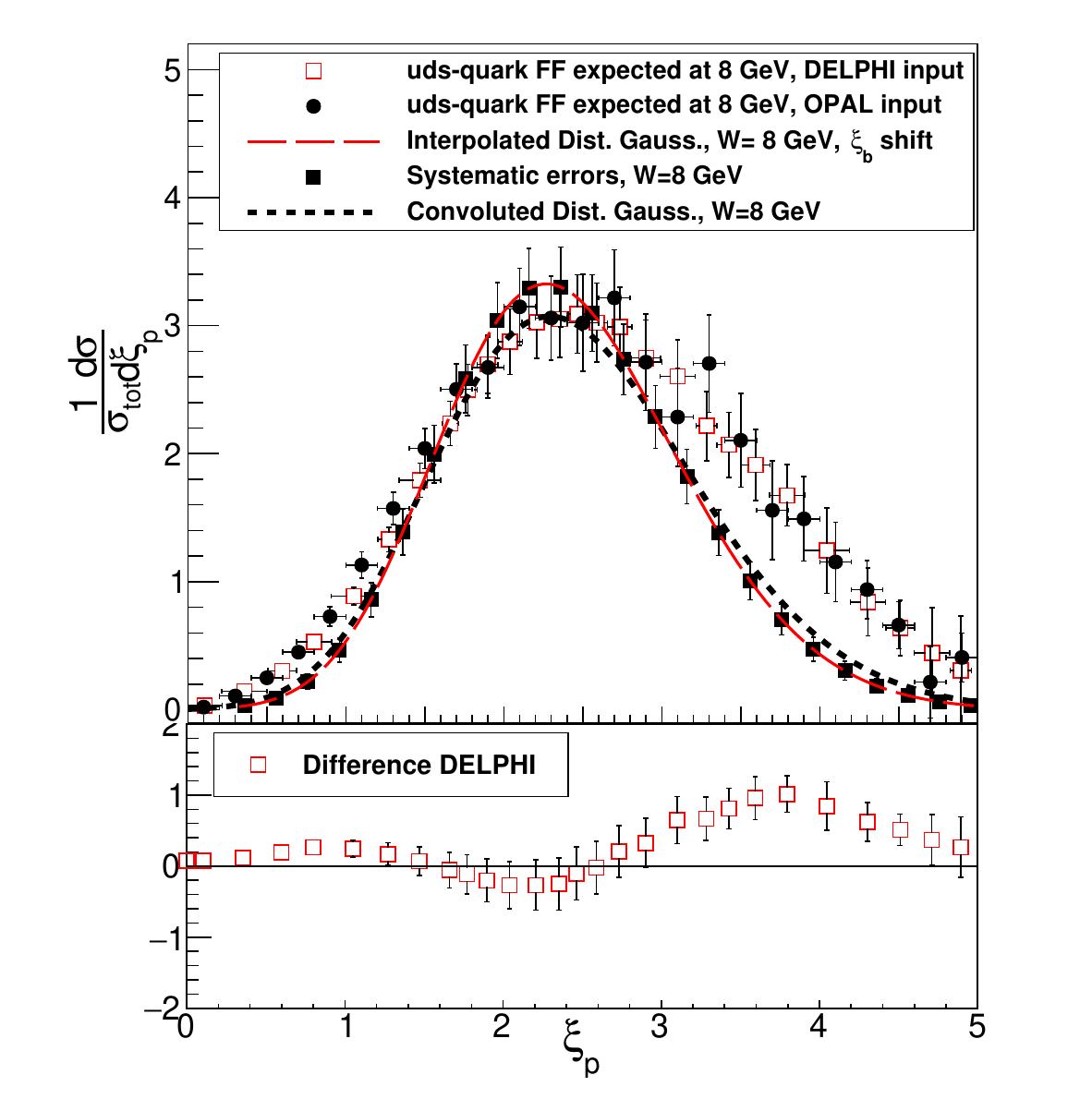}\par 
\hskip -1.0cm
\includegraphics[height=9.5cm,width=9.5cm]{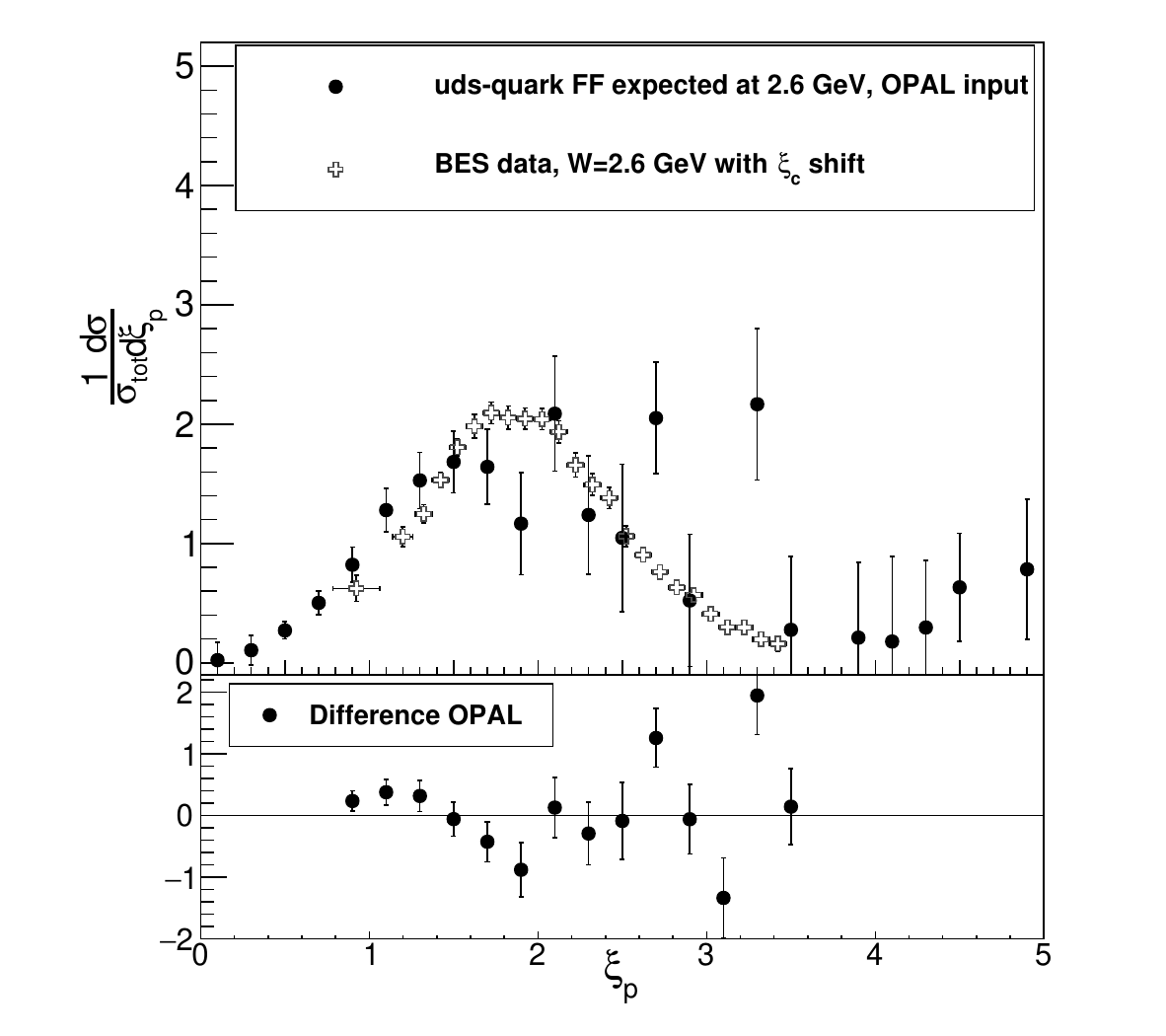}\par 
\end{multicols}
\vskip -0.5cm
\caption{Expected $\xi_p$-Fragmentation Function (FF) at $W_0=8.0$~GeV constructed according to MLLA eq.~\eqref{MLLAeqxi} as difference of $\xi_p$-distributions for uds-quark and b-quark jets with DELPHI and OPAL data as input.}
\label{fig:convolution}
\end{figure*}
These differences are shown in Fig.~\ref{fig:convolution} for the b-quark (left panel) and c-quark fragmentation (right panel). The MLLA expected $\xi_p$-distributions are compared with the experimental $\xi_p$-distribution: for the b-quark at 8~GeV with the distribution obtained in \cite{Kluth:2023umf} by interpolation of the distorted Gaussian, for the c-quark with the observed distribution at 2.6~GeV by the BES collaboration. The differences between the MLLA expected and experimental $\xi_p$-distributions are shown in the lower part of Fig.~\ref{fig:convolution}. For comparison with limiting spectrum curves, the reader is reported to the main reference \cite{Kluth:2023umf}. 

This figure clearly shows that the difference between heavy and light quark fragmentation, in its main features, can just be related to the low energy ``hump backed plateau'' which changes with the MLLA mass scale $W_0= \sqrt{e} M_Q$. When the energy $W_0$ is increased from 2.6 to 8.0~GeV, the $\xi_p$-distribution shifts to a higher mean value $\bar \xi_p$ with larger width and increasing height in agreement with the behaviour known from experiment in absolute terms. This result not only explains the limits of their ratios in Fig.~\ref{fig:exp-deadcone_ratios} for small and large $\xi_p$ with $R\sim 0$ and $R=1$, but also the behaviour in between. 

For the b-quark, the interpolated distribution at 8~GeV approaches quite closely the data in the central region $1\lesssim \xi_p \lesssim 3$, but falls somewhat below the expectations for the very small $\xi_p\lesssim 1$ ($x_p\gtrsim 0.4$) and there is a considerable and very significant excess over the expectation in the large $\xi_p\gtrsim 3$ region. For the c-quark fragmentation there is a good agreement but errors become large for the larger $\xi_p$. 

It appears that the width of the observed distribution in b-quark fragmentation is larger than the expected one. One possible explanation could be that the momentum fluctuations of the heavy quark are larger than anticipated in the MLLA formula eq.~\eqref{MLLAeq} where a fixed energy loss $\langle x_Q\rangle$ is assumed.

The MLLA prediction for b-quark fragmentation and the experimental spectrum at 8~GeV are compared separately for the regions below and above $\xi_p = 3$. To this end, the difference between the expected and experimental distributions at 8~GeV, see Fig.~\ref{fig:convolution}, is fitted to a polynomial function, and the respective multiplicities $\Delta N=N(8\ \rm{GeV})^{MLLA}$ - N(8~GeV)$^{\rm exp}$ are calculated from the integrals over the two $\xi_p$ regions for Distorted Gaussian (Fig.~\ref{fig:convolution}) and the Limiting Spectrum distributions~\cite{Kluth:2023umf}. The results are shown in Tab.~\ref{tab:parameters}.
\begin{figure*}[htpb!]
\begin{multicols}{2}
\hskip -1.1cm
\includegraphics[height=8.0cm,width=9.5cm]{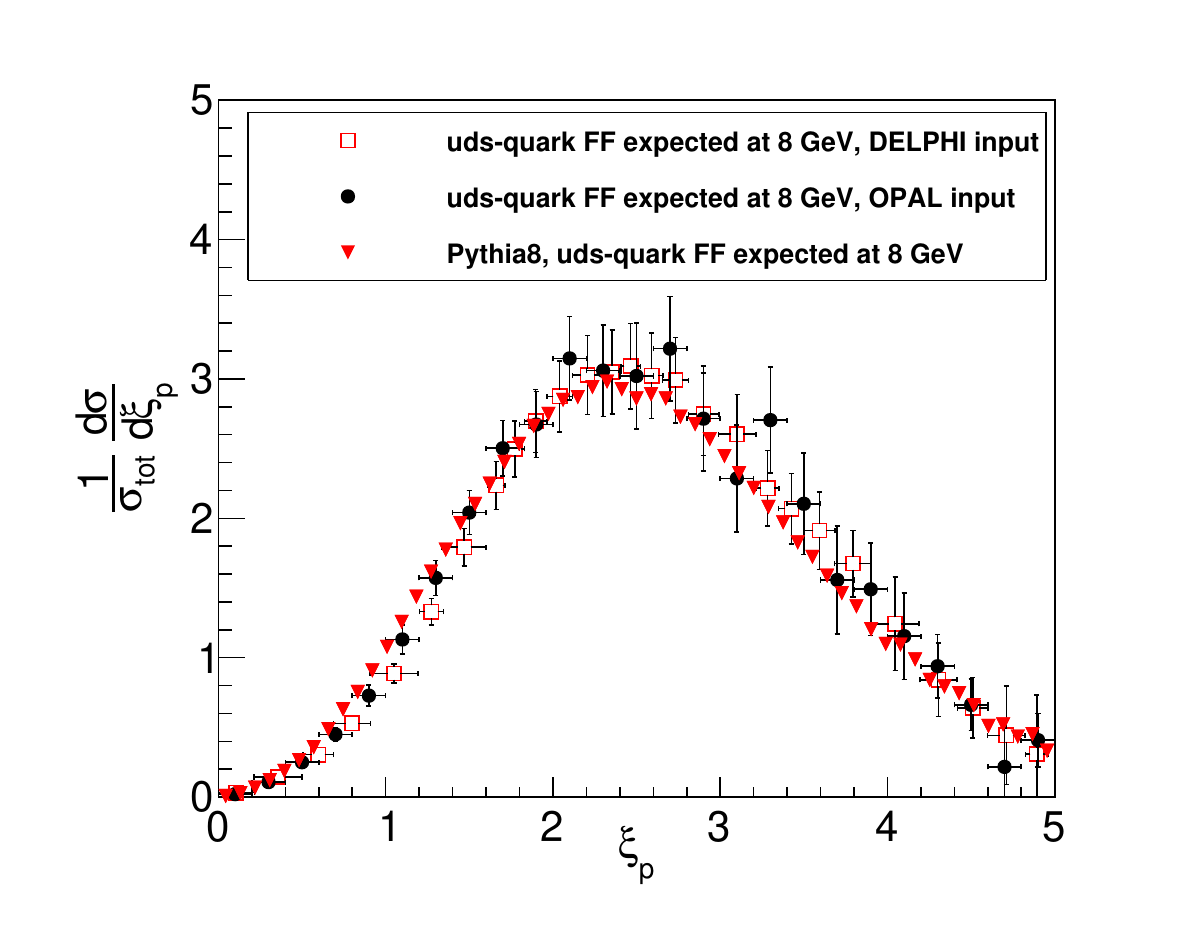}\par 
\hskip -1cm    
\includegraphics[height=8.0cm,width=9.5cm]{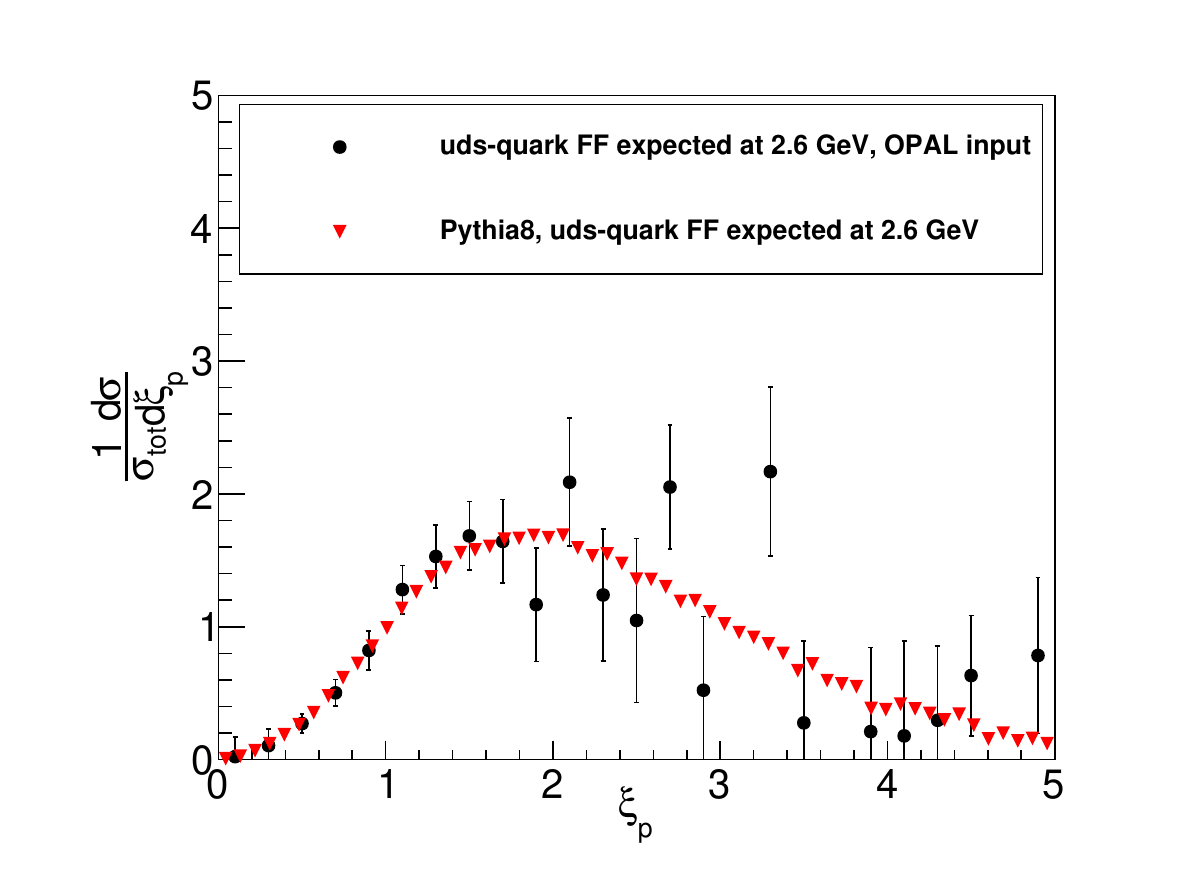}\par 
\end{multicols}
\vskip -0.5cm
\caption{Expected $\xi_p$-Fragmentation Function (FF) at $W_0=8.0$~GeV constructed according to MLLA eq.~\eqref{MLLAeqxi} as difference of $\xi_p$-distributions uds-quark and b-quark jets obtained with Pythia8, DELPHI and OPAL data as input.}
\label{fig:Expectes_HBP}
\end{figure*}

\begin{table}[!bhtp]
\setlength{\tabcolsep}{2.5pc}
\begin{tabular}{cc}
$\xi_p$-range &  Distorted Gaussian \\
$\rm{all} \ \xi_p$ & $\Delta N=1.52\pm0.25$  \\
$\xi_p<3$ & $\Delta N_{\rm low}=0.14\pm0.16$  \\
$\xi_p>3$ & $\Delta N_{\rm high}=1.37\pm0.16$  \\
\end{tabular}
\caption{Integrals $\Delta N$ over the difference data shown in Fig.~\ref{fig:convolution}  between expected and experimental $\xi_p$-distributions for b-quark fragmentation for different $\xi_p$ regions and two interpolating functions at 8~GeV.}
\label{tab:parameters}
\end{table}
$\Delta N$ in Tab.~\ref{tab:parameters} is obtained by summing the data points over the full $\xi_p$ range and it should agree with those published total multiplicities~\cite{Dokshitzer:2005ri}. For DELPHI, one has N(8~GeV)$^{\rm MLLA}$ = 7.87$\pm0.54$ and with $N_{uds}(8\ \rm {GeV})^{exp}$ = $6.1\pm0.3$, one finds $\Delta$N =$1.8\pm 0.6$, which compares well with the value from our fit $\Delta$N = $1.52\pm 0.25$. 

Finally, in figure \ref{fig:Expectes_HBP}, we compare the uds-quark FF at low energy $W_0$ expected from data as shown in Fig. 3 also with the predictions from Pythia8 MCEG. The agreement with DELPHI and OPAL data is very good, in particular, the kinematic region $\xi_p>3$ is well fitted as opposed to the MLLA theoretical prediction in Fig. 3. So the MLLA picture appears to be not complete in this region. This may arise from neglecting multiple gluon emission by the primary heavy quark, also neglected for multiplicity calculations where only the single emission is considered ~\cite{Schumm:1992xt}. This problem deserves further studies. 

\section{Conclusions}

The dead cone effect predicted by perturbative QCD has been studied using data taken at LEP on identified heavy b- and c-quark and light uds-quark fragmentation. 
The dead cone effect for particle production at small angles to the primary quark is also reflected in the production of large momenta, as is typical for the jet structure. In the present study, QCD expectations for the momentum spectra in heavy quark jets based on the MLLA~\cite{Dokshitzer:1991fc,Dokshitzer:1991fd} are investigated. 

At first, we reconstruct the inclusive distributions of charged particle momenta using the variable $\xi_p=\ln(1/x_p)$ in b-quark and c-quark events by correcting for B-hadron and Charm-hadron decays. In the comparison of heavy and light quark fragmentation we observe a convergence of the spectra for large $\xi_p$ ($x_p\to 0$) but a strong suppression of the fragmentation functions of the heavy b- and c-quarks with respect to the one of the light uds-quarks with decreasing $\xi_p\to 0$ (increasing $x\to 1$) down to a fraction of $\lesssim 1/10$. This observed almost complete suppression reflects the presence of the dead cone with a high significance ($\gg 5\sigma$). There is a characteristic difference between b- and c-quark fragmentation, in that the decrease for the c-quark is shifted towards lower $\xi_p$ as compared to the b-quark. It would be desirable to replace the MCEG based subtraction of the charged heavy hadron decay products by an experimental measurement in order to remove any residual model dependence.

The $\xi_p$-distributions derived from experimental data are then compared directly with the QCD expectations within the MLLA following the hypothesis of Local Parton Hadron Duality (LPHD). This QCD analysis provides a quantitative explanation of the dead cone effect: the difference between the heavy and light quark fragmentation functions in the variable $\xi_p$ at high c.m.s.\ energy $W$ is just given by the $\xi_p$-fragmentation function at the lower energy $W_0=\sqrt{e}M_Q$ with the heavy quark mass $M_Q$ (see eq.~\eqref{MLLAeq}).

The equation~\ref{MLLAeq}, an estimate within MLLA, is tested first with the experimentally observed or derived $\xi_p$-distributions as input. For both the b-quark and the c-quark fragmentation this equation is found to be well supported in the central kinematic region around the peak of the $\xi_p$-spectrum at scale $M_Q$ corresponding to the momentum range $x\lesssim 0.4$ and $p\gtrsim \Lambda$ with the QCD scale $\Lambda\sim 250$ MeV. It explains quantitatively the suppression of both fragmentation functions down to about $1/10$ of the one for uds-quarks. The different suppression profiles of c- and b-quark fragmentation are directly related to the different shapes of the $\xi_p$ spectra (the ``hump-backed plateau'') at the respective c- and b-quark mass scales $W_0$, i.e.\ at the c.m.s.\ energies 2.7 and 8.0~GeV respectively.
In the MLLA estimate eq.~\eqref{MLLAeq}, the mean fractional momentum $\langle x_Q \rangle$ of the heavy quark appears as additional (known) parameter. This parameter is important for the successful quantitative description. The interplay between heavy quark fragmentation and energy loss deserves further attention.

In the kinematic region of small $\xi_p\lesssim 1$ ($x_p\gtrsim 0.4$), at low values for the heavy quark fragmentation functions, violations of the MLLA relation have been observed. This is to be expected in the present approximate scheme using a shifted spectrum at the low energy $W_0$. Integrating the fragmentation function of the b-quark
over the important region $\xi_p\lesssim 3$ ($p\gtrsim \Lambda$) yields the respective multiplicity which is found in good agreement with the MLLA expectation. 
A large and significant excess of particle production in b-quark fragmentation over these expectations is observed for large $\xi_p\gtrsim 3$ which concerns the region of very soft particle production with $p\lesssim \Lambda$. 
The total excess multiplicity in this kinematic region corresponds quantitatively to the excess over MLLA expectations already noted in the previous study of the full multiplicity~\cite{Dokshitzer:2005ri}. This discrepancy comes from a kinematic region outside the validity of the perturbative approach. In case of the c-quark fragmentation 
no such excess at large $\xi_p$ can be resolved within the larger errors and there is a satisfactory agreement between prediction and experimental data for not too large momenta ($\xi_p\gtrsim 1$). The Pythia8 MCEG describes the $\xi_p$-spectra rather well over the full kinematic range observed for both the c- and the b-quark events.

 Finally, our results could be of direct relevance to heavy quark tagging techniques. Traditional heavy quark jet tagging algorithms only use variables derived from particles associated with the heavy hadron decay, see e.g.~\cite{Barker:2010pva} for a review of algorithms used at LEP. Recent developments using advanced machine learning techniques (see e.g.~\cite{Qu:2019gqs,ATLAS:2022rkn} and references therein) include all objects associated with the jet and thus their improved performance compared to traditional algorithms could be related at least partially to the dead cone effect.
 
\bibliographystyle{unsrt}
\bibliography{hbpdeadcone}
\end{document}